\documentclass[aps,prl,reprint,superscriptaddress,longbibliography]{revtex4-2} %

\usepackage[T1]{fontenc} 
\usepackage[latin9]{inputenc} 
\usepackage{graphicx}

\usepackage{amsmath}  
\usepackage{gensymb} 
\usepackage{epstopdf} 
\usepackage{setspace}  
\usepackage{datetime} 
\usepackage{dcolumn} 
\usepackage{comment}
\excludecomment{skipover} 

\usepackage{color}
   
  \definecolor{dgreen}{rgb}{0.00, 0.5, 0.00}
  
\usepackage{comment}
\excludecomment{skipover} 
\usepackage{natbib, hyperref}
\hypersetup{colorlinks=true,linkcolor=blue, filecolor=blue,urlcolor=blue,citecolor=blue,}

\begin{document}
\title{No band gap, no problem:
Defects in InAs using a band-avoiding occupation-constrained density functional theory}
\author{Peter A. Schultz}
\email{paschul.xnl@gmail.com}
\affiliation{Sandia National Laboratories,
Albuquerque, NM~87185, USA}
\author{Arthur H. Edwards}
\affiliation{Air Force Research Laboratory,
Kirtland Air Force Base, NM 87117, USA}
\author{Evan M. Anderson}
\affiliation{Sandia National Laboratories,
Albuquerque, NM~87185, USA}
\author{Anthony C. Knighton}
\affiliation{Air Force Research Laboratory,
Kirtland Air Force Base, NM 87117, USA}
\author{Leopoldo Diaz}
\email{lndiaz@sandia.com}
\affiliation{Sandia National Laboratories,
Albuquerque, NM~87185, USA}
%
%
%
%
%
\begin{abstract}
Density functional theory (DFT) underestimates the experimental band gap---%
the infamous band gap problem. As the band gap defines the energy scale of defect levels, this complicates computation of charge transition energies for atomic defects. In the extreme case of narrow-gap semiconductors, the DFT band gap collapses to zero, seemingly precluding quantitative predictions of defect levels. We present a band-avoiding occupation-constrained DFT (ba-occ-DFT) approach that prevents spurious occupation of band-edge states and enables reliable total energy calculations of atomic defects. Application to indium arsenide (InAs) shows that ba-occ-DFT circumvents the band gap problem, separates band-edge errors from defect level calculations, and enables rigorous defect level predictions in a narrow-gap semiconductor despite a zero DFT band gap.
\end{abstract}
\maketitle
%

%
%



%


Density functional theory (DFT)~\cite{HK64,KohnSham} within the local density approximation (LDA)
or generalized gradient approximation (GGA) is widely used with great success
for structural properties of bulk crystals and defects in first-principles material simulations,
with typical accuracy of a couple tenths of an electron Volt (eV) over a wide range of applications. This success in structural energetics contrasts sharply with DFT's inadequate 
description of the fundamental band gap in semiconductors and insulators.
The Kohn-Sham (KS) band gap, the difference between computed KS eigenvalues
for the highest occupied valence band edge (VBE) state and the first unoccupied conduction band edge (CBE) state,
greatly underestimates the experimental gap~\cite{DFTGAP1,DFTGAP2,DFTGAP2b}---%
the infamous ``band gap problem''.
The band gap problem reaches its extreme in narrow-gap semiconductors such as indium arsenide (InAs),
where the CBE collapses to the VBE at the $\Gamma$-point in the Brillouin Zone (BZ), 
which seemingly precludes accurate predictions of defect levels (charge state transition energies) using DFT.
%
In this Letter, we separate the band gap problem from the problem of calculating localized defect states.
Using InAs as the exemplar,
we demonstrate a band-avoiding occupation-constrained DFT (ba-occ-DFT) for predictive calculations of defect levels,
even in semiconductors with a zero DFT gap.

Indium arsenide is a zinc-blende structure semiconductor with a 0.42~eV band gap (at 0K)~\cite{InAs-Eg1,InAs-Eg2,InAs-Eg3}
and high electron mobility,
an exemplar of a family of narrow band gap III-V alloys \{Al,Ga,In\}\{P,As,Sb\}
with great potential for low-power high-speed electronics and infrared applications.
Atomic displacement damage degrades electronic performance of detector materials
in the harsh radiation environments of space,
sparking focused efforts to determine the nature of the atomic-scale defects responsible
and thereby engineer effective mitigation strategies.
The mainstay of defect characterization in semiconductors,
deep-level transient spectroscopy (DLTS),
has been impaired in narrow-gap systems by leakage currents mediated by surface effects~\cite{Nelson20,Rigo26}.
Simulations of defects in InAs using DFT~\cite{Lee96,Hoglund06,Tahini13,Chroneos14,Reveil17,Miao19,Peng19,Brennaman23}
are also impaired, 
by the band gap problem and by finite-size errors.
No consensus theory has emerged, and no two previous studies appear to agree on almost any defect physics in InAs.

Contrary to conventional wisdom concerning the band gap problem and defect levels, we previously demonstrated that a total-energy-based DFT method for charged defects~\cite{PAS06} predicted defect levels in silicon (Si) to within 0.1(average)-0.2(max)~eV of experiment for a wide variety of defects
across a full experimental band gap, despite a band gap problem.
The crucial feature was incorporating rigorous Coulomb boundary conditions appropriate for charged supercells
via the local moment countercharge (LMCC) approach~\cite{PAS99,PAS00}.
Hybrid functional approaches~\cite{HSE03,HSE06}, despite fixing the band gap problem,
have not (to date) replicated this broad validation benchmark in Si.
Similarly, for gallium arsenide (GaAs), DFT+LMCC calculations reliably predict defect levels to 0.1~eV accuracy~\cite{PAS09,PAS-E1E2,PAS-E3c},
accuracy sufficient to populate predictive defect chemical kinetics models of defect evolution in irradiated GaAs~\cite{Diaz25}.
Even the most recent comprehensive Heyd-Scuseria-Ernzerhof HSE06$+$jellium hybrid functional~\cite{HSE03,HSE06} defect study~\cite{Fluckey26}
failed to predict a single viable defect candidate for any radiation-induced defect level in GaAs~\cite{Bourg88}.


\begin{figure}[h]
\includegraphics[trim={0.6in 1.8in 3.6in 1.0in}, clip, width=\columnwidth]{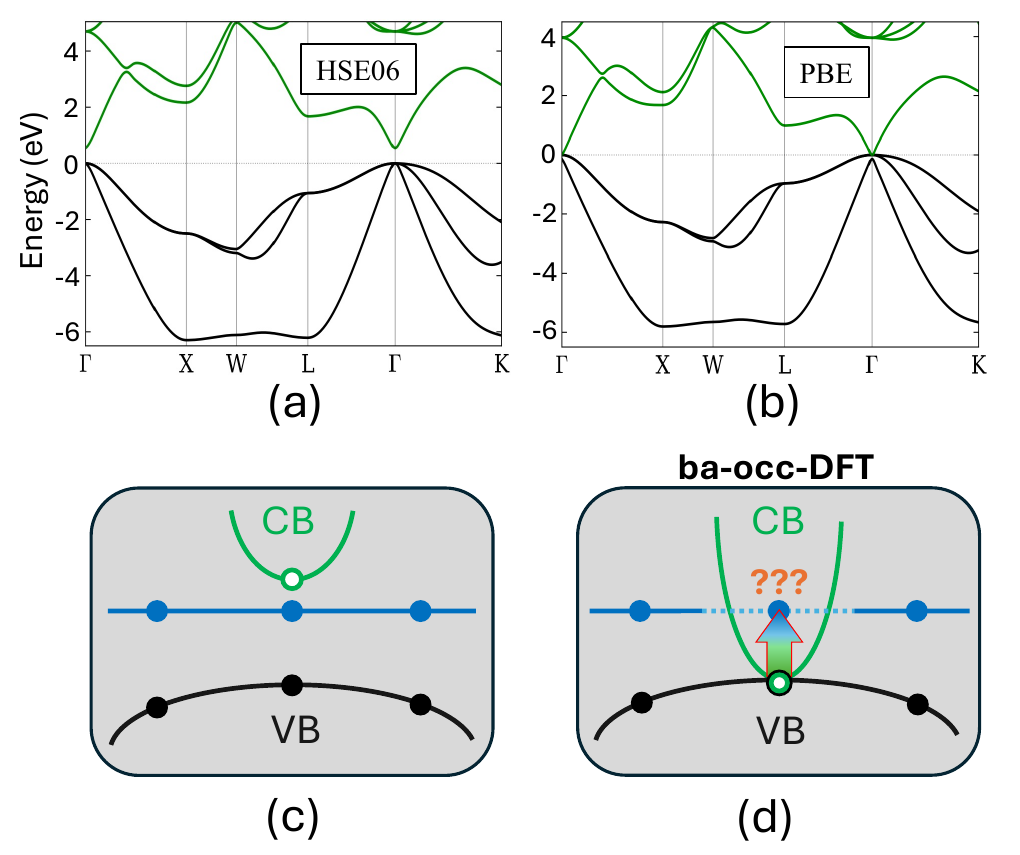}
\caption{
\label{FIG:InAs-BS}
Band structures for crystalline InAs and idealized defect levels:
(a) InAs HSE06 band structure using VASP;
(b) InAs PBE band structure using SeqQuest;
(c) Idealized defect level (blue)---flat, dispersionless state in the band gap;
(d) The band-avoiding occupation-constrained DFT: ``exciting'' the electron at $\Gamma$ from the spurious CBE to the defect state.
 }
\end{figure}
Figure~\ref{FIG:InAs-BS} illustrates the central issue addressed in this Letter. Our HSE06 band structure of crystalline InAs (computed {\emph{without}} spin-orbit coupling) has a direct band gap of 0.54~eV at the $\Gamma$-point [Fig.~\ref{FIG:InAs-BS}(a)], in quantitative agreement with previous results~\cite{Kim09}. In contrast, PBE collapses this gap at $\Gamma$ [Fig.~\ref{FIG:InAs-BS}(b)]. As illustrated in Fig.~\ref{FIG:InAs-BS}(c), an ideal localized defect level is dispersionless (flat) across the Brillouin Zone (BZ) and cuts through the gap between the CBE and VBE. Supercell calculations of defects using $k$-point grids offset from $\Gamma$ show defect states that are cleanly distinct from the off-$\Gamma$ band-edge states~\cite{PAS09}. The crucial question then becomes: what happens at $\Gamma$, and what does that imply for computing a defect level?

Our central hypothesis, depicted in Fig.~\ref{FIG:InAs-BS}(d), is that the localized defect state continues through $\Gamma$ above the spuriously descending CBE state and by selectively occupying the localized defect state at $\Gamma$ and avoiding occupation of the CBE state, we can compute the total energy of the localized defect accurately uncorrupted by band-edge states.
Defect states away from $\Gamma$, already in the KS gap, are sampled with standard Aufbau occupations, while the states at $\Gamma$ require switching occupation from the CBE state to the defect state.

The standard Aufbau occupation scheme in ground-state DFT occupies electron eigenstates from lowest to highest energy until no electrons remain. Occupation-constrained DFT modifies this procedure by imposing non-Aufbau occupations, skipping selected lower-energy states and occupying selected higher-energy states across the BZ, enabling the modeling of excited states~\cite{PAS-occdft}. Here, we adapt this occ-DFT machinery to impose $\Gamma$-specific occupation constraints, skipping occupation of the spurious CBE state and instead occupying the localized defect state to compute a clean defect ground state.

We apply this ba-occ-DFT approach to simple intrinsic point defects in InAs: antisites (As$_\text{In}$ and In$_\text{As}$), single vacancies ($v_\text{In}$, $v_\text{As}$), the divacancy ({$vv$}),
and interstitials (As$_i$, In$_i$). The defect calculations using the PBE functional~\cite{GGA-PBE}
are performed with the {\sc SeqQuest} code~\cite{SEQQ} in 64, 216, 512, and 1000-atom supercells,
$N{\times}N{\times}N$ expansions ($N=2{-}5$) of the conventional 8-atom cubic cell.
Regular $2{\times}2{\times}2$ $k$-grids to sample the BZ are required to converge the supercell calculations ($3{\times}3{\times}3$ for the 64-atom cell).
The lattice parameter is fixed to the computed PBE lattice parameter, $a_0=6.19$~\AA. During structural relaxation, we explored possible symmetry-lowering distortions, especially site-shifts, and achieved forces less than 0.0002~Ry/Bohr ($\sim$5~meV/\AA).
The LMCC method enforces rigorous boundary conditions
for charged defect calculations~\cite{PAS99,PAS00},
to remove finite-size effects~\cite{PAS06}.
We also perform additional HSE06 hybrid functional calculations for perfect crystal supercells and the arsenic antisite using VASP~\cite{VASP-code,VASP-PAW},
for small, computationally tractable 64 and 216-atom supercells fixed to the experimental lattice parameter
($a_0=6.06$~\AA)~\cite{Nahory78}.
Additional details are described in the End Matter (EM) and analysis in the Supplemental Material (SM)~\cite{SUPPL}.  

\begin{figure}[h]
\includegraphics[trim={0.5in 1.6in 4.6in 1.6in}, clip, width=\columnwidth]{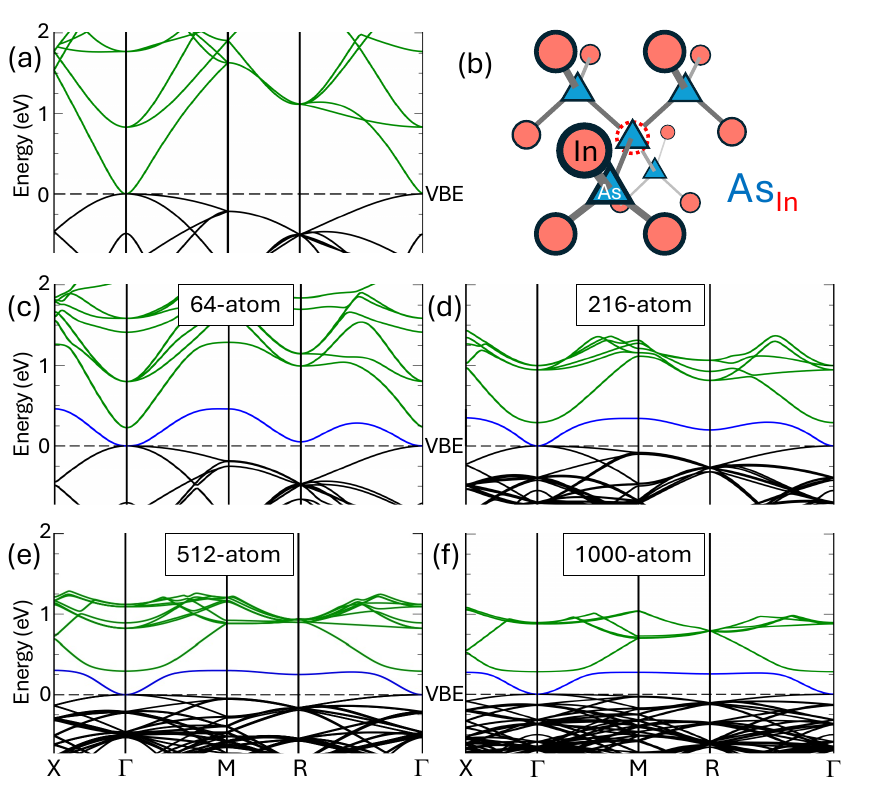}
\caption{
\label{FIG:aAs0-BS}
PBE band structure (a) the perfect crystal 64-atom supercell and As$_\text{In}$(0): (b) structure and band structure in the (c) 64-, (d) 216-, (e) 512-, and (f) 1000-atom supercells. The smaller supercells show large defect state dispersion and strong hybridization with the CBE state,
illustrating the perils of using small supercells (or inadequate $k$-point samples).
 }
\end{figure}

The As antisite, the canonical defect in GaAs, is associated with the $EL2$ defect trap~\cite{Bourg88,EL2-1,EL2-2} observed in as-grown (As-rich) GaAs. Figure~\ref{FIG:aAs0-BS} plots the band structures for the analogous arsenic antisite expected in InAs.
In the smallest 64-atom supercell, the antisite seemingly opens a gap at $\Gamma$
and introduces a defect state in the gap across the BZ, touching the VBE at $\Gamma$.
However, observe:
(1) this presumed defect band disperses by 0.5~eV---larger than the band gap!---%
and (2) the conduction band structure [Fig.~\ref{FIG:aAs0-BS}(c)] is hugely altered from the perfect crystal [Fig.~\ref{FIG:aAs0-BS}(a)]. The 216-atom supercell still shows significant dispersion, but the defect state progressively flattens in the 512-atom and 1000-atom supercells. 

Using the hypothesized switched CBE-defect state order at $\Gamma$,
the As$_\text{In}$(0) defect state dispersion remains large in the 64-atom supercell, 0.45~eV (at the special points). This dispersion narrows to 0.15, 0.05, and 0.02~eV in the 216-, 512-, and 1000-atom supercells, respectively---asymptotically satisfying the requirement of a flat (localized) defect state distinct from the descending CBE state.
It is the (spurious) CBE state that touches the VBE.
The defect band and CBE switch character through an avoided crossing between $\Gamma$ and the BZ boundary.


The band gap problem has motivated the use of hybrid functional approaches~\cite{HSE03,HSE06} for defect calculations~\cite{RMP14} where the fraction of explicit Hartree-Fock exchange relative to DFT exchange can be empirically tuned to match the experimental gap. This raises the obvious question: why not just use the conventional HSE methods~\cite{RMP14}? To date, there have been no published hybrid functional calculations of defects in InAs or comparable narrow-gap III{-}V.

\begin{figure}[h]
\includegraphics[trim={1.4in 1.6in 1.0in 1.6in}, clip, width=\columnwidth]{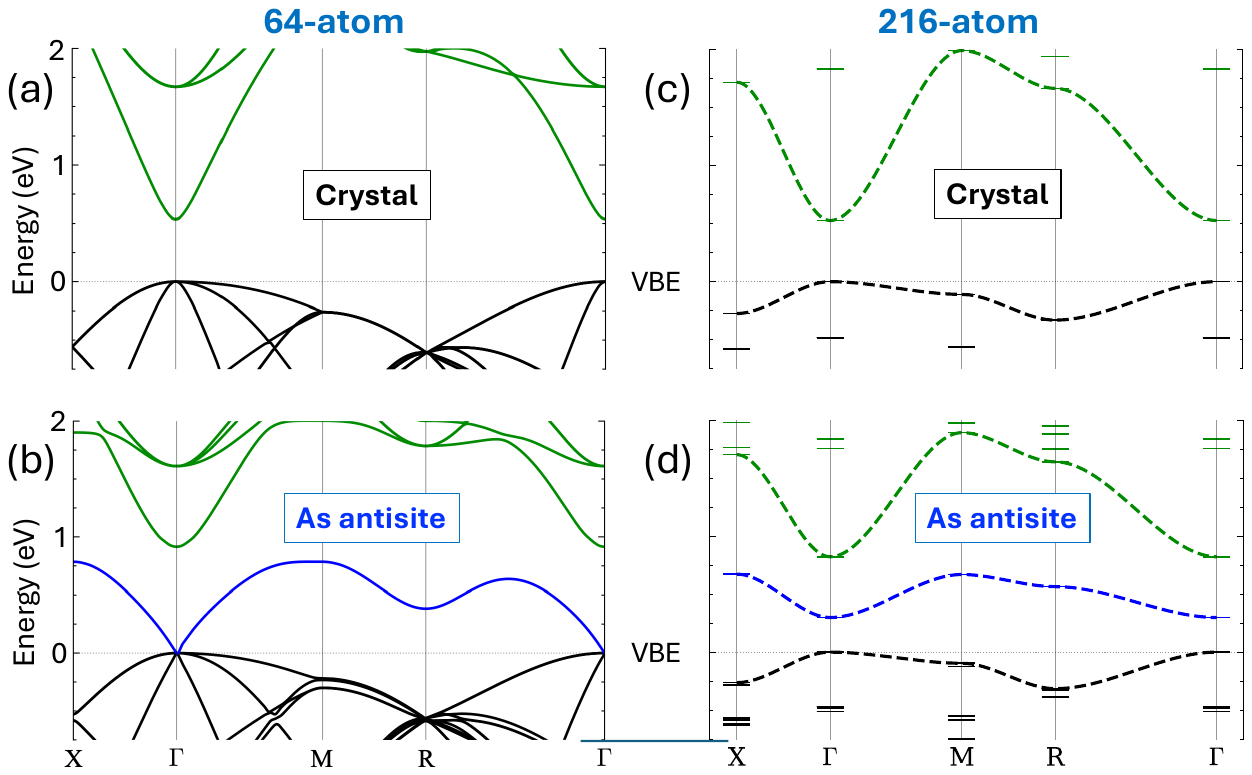}
\caption{
\label{FIG:aAs0-HSE}
The HSE06 band structure:
(a) for the perfect crystal and (b) As$_\text{In}$(0) in the 64-atom supercell;
(c) at special points in the perfect crystal and (d) As$_\text{In}$(0) in the 216-atom supercell. Dashed lines guide the eye across the plot; the full HSE06 band structures were too computationally costly.
}
\end{figure}

The antisite band structures in Fig.~\ref{FIG:aAs0-HSE}
show that the hybrid functional HSE06 does not improve upon PBE.
Our HSE06 64-atom As$_\text{In}$(0) band structure in Fig.~\ref{FIG:aAs0-HSE}(b)
closely mimics our PBE result in Fig.~\ref{FIG:aAs0-BS}(c):
a defect band with $\approx$0.5~eV dispersion (the HSE06 band gap!) apparently descends to the VBE at $\Gamma$.
The HSE06 216-atom defect band still disperses $>$0.3 eV
(by inspection at the special $k$-points),
more than with PBE [Fig.~\ref{FIG:aAs0-BS}(d)],
and the apparent CBE rises to 0.9~eV.
The HSE06 $\Gamma$-point levels are entirely untrustworthy in these smaller supercells. HSE06 opens the crystal band gap, but opening the band gap alone does not guarantee clean localized defect states in the gap using finite supercells.

We see the same flattening of defect bands across the BZ for all InAs defects in the PBE calculations.
This result indicates that a ba-occ-DFT, which skips occupation of the spurious CBE state at $\Gamma$, is plausible.
Our ba-occ-DFT implementation indeed leads to stable self-consistent calculations.
The hypothesized localized defect character (and convergence with respect to $k$-sampling)
is confirmed in a detailed assessment of the As$_\text{In}$ calculations
as a function of defect charge, supercell size, $k$-point sampling, and occupation method (see SM~\cite{SUPPL}).
The As$_\text{In}$ has stable neutral, and (1+) and (2+) donor states.
The ground states retain the $T_d$ symmetry, similar to the As antisite in GaAs~\cite{PAS09}.
Charge transitions cause reductions in the As--As bond length,
from 2.651(3) in the neutral, to 2.611(1) in the (1+), and 2.549~{\AA} in the (2+) charge state.
Hence, the bond length can be used to infer the charge state.
Using the off-$\Gamma$ $k$-grids, the defect states lie cleanly in the gap at all sampled $k$-points.
The $\Gamma$-$(2{\times}2{\times}2)$ ba-occ-DFT results yield formation energies comparable to (no more than 30-40~meV larger than)
the off-$\Gamma$ grids.
Comparison of the off- and on-$\Gamma$ sampling results cannot
conclusively discriminate the character of the states at $\Gamma$. Only small differences are expected because $\Gamma$ is weighted only~$\frac{1}{8}$.

The inverted antisite-CBE character is unambiguously demonstrated in $\Gamma$-only calculations (see the SM~\cite{SUPPL}).
The $\Gamma$-only ba-occ-DFT
results largely agree with the $(2{\times}2{\times}2)$ results.
However, the $\Gamma$-only standard Aufbau fails for (0) and (1+):
the As--As bond distance in all charge states collapses to 2.55~{\AA},
indicating an As$_\text{In}$(2+) plus conduction band electrons.
The As$_\text{in}$(2+) state itself has no {\emph{occupied}} gap states and therefore does not suffer from a spuriously occupied CBE state.

We apply this ba-occ-DFT 
to survey all simple primary displacement damage intrinsic defects in InAs.
First, we perform standard Aufbau calculations for each defect in the $\Gamma$-centered $k$-grids.
If inspection reveals a spuriously occupied CBE state at $\Gamma$, 
then we apply a ba-occ-DFT to empty the CBE state and occupy defect states.
The total energies and ensuing defect levels are computed with the LMCC-based finite-defect model~\cite{PAS06,PAS-3C-SiC}
(additional details in the SM~\cite{SUPPL}).

\begin{figure}[h]
\includegraphics[trim={0.6in 3.5in 3.3in 1.2in}, clip, width=\columnwidth]{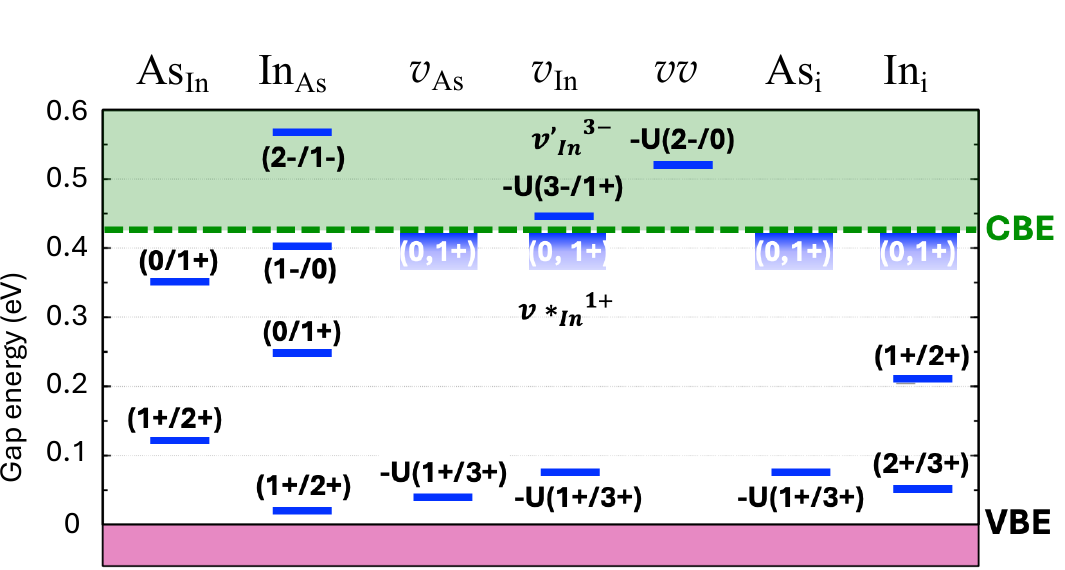} 
\caption{
\label{FIG:InAsLevels}
Predicted InAs defect level diagram,
using the ba-occ-DFT in 1000-atom supercells.
The VBE is bounded by the In$_\text{As}(2+/1+)$,
the CBE is set by the experimental gap.
}
\end{figure}

Figure~\ref{FIG:InAsLevels} shows the predicted ba-occ-DFT defect level diagram from the $\Gamma$-$(2{\times}2{\times}2)$ 1000-atom supercell. These defect levels are well converged (to within $\approx$20~meV) with the 216-atom supercell,
despite defect band dispersions spanning more than 0.2~eV---half the band gap~\cite{SUPPL}.
This demonstrates that the ba-occ-DFT effectively isolates a clean defect state and highlights the need to use a converged $(2{\times}2{\times}2)$ $k$-grid to sample the defect-band dispersion
(rather than only a $\Gamma$-only calculation, for example).

As expected, in analogy to GaAs,
the As antisite has two donor states:
(0/1+) at 0.35~eV and (1+/2+) at 0.12~eV.
The As$_\text{In}(0)$ formation energy is small, only 1.09~eV, and decreases to 0.62~eV for the As$_\text{In}$(2+) in $p$-type InAs (Fermi level at the VBE).
As in GaAs~\cite{EL2-1,EL2-2,PAS09},
the As antisite is a low-energy defect,
that might have appreciable population in as-grown InAs.
The InAs intrinsic defects largely adopt the same structures
as their counterparts in GaAs~\cite{PAS09}
but with a much more limited range of stable charge states---%
fewer charge states, especially negative charge states, are stable in InAs.

In contrast to all previous theory~\cite{Lee96,Hoglund06,Tahini13,Chroneos14,Reveil17,Miao19,Peng19,Brennaman23}, we find that although the simple $v'_\text{In}$(3-) is a localized defect state, $v'_\text{In}$ is not thermodynamically stable in the gap.
A site-shifted $v^*_\text{In}$=As$_\text{In}$--$v_\text{As}$ form proves more stable:
the smaller neighboring As atom hops into the site vacated by the larger In.
The resulting four(4)-electron negative-$U$ $v'_\text{In}$(3-)/$v^*_\text{In}$(1+) transition lies at 0.45~eV---above the experimental band gap.
The site-shift $v^*_\text{In}$ undergoes a second $-U$(1+/3+) transition at 0.07~eV.

Conversely, $v_\text{As}$ is stable only as a simple vacancy; no site-shift configuration is stable.
It is stable in positive charge states,
with a $-U$(1+/3+) transition at 0.04~eV, just barely above the VBE.
The divacancy is stable only as a neutral defect in the gap;
a $-U$ transition to a $(2-)$ state lies above the CBE.

The In$_\text{As}$ cleanly adopts localized charge states from (2+) through (1-);
a (2-) state at 0.55~eV is less certain.
The (2+) charge state, while cleanly discriminated from the VBE (in the largest 1000-atom supercell),
is very shallow, and serves as the upper bound of the VBE.
The (1-) state at 0.41~eV is the highest clean defect state, hence a lower bound on the CBE,
defining an effective defect band gap (EDBG)
in reasonable agreement with experiment.

The interstitial defects take positive charge states from (1+) through (3+),
the As$_i$ in a $-U(1+/3+)$ near the VBE,
and the In$_i$ in two separate single-stepped levels.

Ultimately, experiment provides the most important validation of defect calculations.
In InAs, very little is definitively known about any atomic-scale defect physics
because of the daunting technical challenges of performing defect-discriminating experiments
in narrow-gap semiconductors~\cite{Zachman93,Rigo26}.
By analogy to GaAs, one expects an $EL2${-}like state in as-grown InAs from the As antisite.
Our calculations predict a donor state at 0.35~eV, consistent with states observed by Salman {\it et al.} at CBE$-0.06$~eV~\cite{Salman90} and by Murawski {\it et al.} at CBE$-0.05$~eV~\cite{Murawski24}. Although these associations are consistent with experiment,
they are not conclusive; additional discriminating experiments are needed to confirm the identification.  
For example, our prediction of a second donor state just above the VBE does not (yet) have an experimental counterpart.

More illuminating is the observation in the most recent experimental work for irradiated InAs of a broad ``shoulder'' in DLTS,
interpreted as suggestive of emission from multiple shallow electron defect levels~\cite{Rigo26}.
Our results offer a simple explanation of this unprecedented observation, not observed in the chemically similar GaAs or other III{-}V's:
emission from primary displacement defects.
The highest stable charge state for
either a single vacancy or interstitial is (1+); we find no neutral charge states
However, a charged defect in a semiconductor is guaranteed to trap an effective-mass-like state~\cite{Sok78}; for a positively charged defect, this corresponds to binding an electron and producing a shallow donor state. Such a shallow state cannot be encompassed within the volume of a small finite supercell~\cite{SHALLOWS}. Thus, the termination at a stable (1+) charge state, even in the 1000-atom supercells, predicts that all the single primary defects will be shallow donors (Fig.~\ref{FIG:InAsLevels}). This prediction is consistent with the broad shallow-emission observed experimentally.

Interestingly, the ba-occ-DFT proves only necessary for CBE-crossing defect states;
we find no instances of VBE-crossing defect states.
The problem is not, as often asserted,
that local DFT overly delocalizes defect states.
The defect states computed with PBE are well localized,
as well localized as---or better localized than---those computed with a hybrid functional.
The problem is that DFT overstabilizes delocalized states,
here bringing the CBE to too low an energy.
Our analysis indicates that the consequences of this
unphysical stabilization of a delocalized state (the CBE collapse at $\Gamma$)
is readily avoided with a ba-occ-DFT
for defect calculations in semiconductors.

We have introduced a band-avoiding occ-DFT approach that enables clean total-energy calculations of defect formation energies and defect transition levels in semiconductors with a severe band gap problem.
The PBE+LMCC ba-occ-DFT results are consistent with what limited data exist for atomic defects in InAs,
an extreme case with zero DFT band gap.
The crucial feature is to strictly enforce occupation of local defect states and avoid spurious occupations of band states,
while using an adequate $k$-sampling to converge a strongly dispersing defect state inevitable in a narrow-gap system.
Our analysis indicates that satisfying this criterion can be accomplished using a relatively simple occupation-constrained DFT approach.
The ba-occ-DFT approach can be straightforwardly implemented into any DFT supercell code.
This extends the accuracy of DFT+LMCC into narrow-gap systems, where conventional hybrid$+$jellium approaches are computationally impractical and have failed to provide comparable accuracy.

{\it Acknowledgments---}%
Sandia National Laboratories is a multimission laboratory managed and operated by National Technology and Engineering Solutions of Sandia, LLC.,
a wholly owned subsidiary of Honeywell International, Inc., for the U.S. Department of Energy's National Nuclear Security Administration under contract DE-NA0003525.
This work (SAND2026-23880O) was partially supported by Laboratory Directed Research and Development (LDRD) projects (No. 242427 and No. 233119).
This paper describes objective technical results and analysis.
Any subjective views or opinions that might be expressed in the paper do not necessarily represent the views of
the U.S. Department of Energy, or the official policy or position of
the Department of the Air Force, the Department of Defense, or
the United States Government. Approved for public release: distribution is unlimited.
We thank the AFRL DSRC and Navy DSRC for providing high-performance computing resources used in this work. Approved for public release; distribution is unlimited. Public Affairs release approval number: AFRL-2026-3355.
\begin{skipover}
This paper is coauthored by employees of 
National Technology \& Engineering Solutions of Sandia, LLC under Contract No. DE-NA0003525 with the U.S. Department of Energy (DOE).
The employees own all right, title and interest in and to the article and are solely responsible for its contents.
The United States Government retains and the publisher, by accepting the article for publication, acknowledges that the United States Government retains a non-exclusive, paid-up, irrevocable, world-wide license to publish or reproduce the published form of this article or allow others to do so, for United States Government purposes.
The DOE will provide public access to these results of federally sponsored research in accordance with the DOE Public Access Plan https://www.energy.gov/downloads/doe-public-access-plan.
\end{skipover}

\bibliography{aps}

\section{End Matter}


\noindent 

In the {\sc SeqQuest} calculations, we use a $Z{=}3$ In pseudopotential (PP) with the semicore $3d^{10}$ electrons in the core and a $Z{=}5$ $s^2p^3$ PP for As.
Both pseudopotentials include nonlinear core corrections to ensure optimal transferability. This large-core ``$d0$'' In pseudopotential performs as well as the ``$d10$'' small-core pseudopotential for calculations of InAs bulk crystal properties~\cite{SUPPL}, justifying the use of the more computationally economical $d0$ PP for In. We employ double-zeta plus polarization atom-centered basis sets built from contracted Gaussian functions, augmented by "floating" orbitals at vacancy sites to minimize basis errors.

For the Aufbau occupations, we use a Discrete Defect Occupation scheme that imposes
uniform occupations across all $k$-points.
Each band has the same occupations everywhere in the Brillouin Zone,
such that the defect occupations map asymptotically
to the ideal isolated defect occupations
in the limit of large supercells. 

Metallic inter-$k$ movement of electrons is not allowed.
The occ-DFT-based state occupation changes are referenced to this clean baseline. 
The (artificial) electronic temperature is strictly zero in all SeqQuest calculations,
preventing partial (or negative) occupations of orbitals.

The standard HSE06 hybrid functional (explicit exchange mixing factor 0.25) calculations are all performed using VASP~\cite{VASP-code,VASP-PAW}.
Regular $\Gamma$-centered $k$-grids of $6\times6\times6$ and $2\times2\times2$ were used for the FCC primitive cells and supercells, respectively.
Our VASP calculations used a plane-wave basis set with an energy cutoff of 400~eV.

\newpage

\begin{figure} [h!]
    \centering
\includegraphics[width=\textwidth]{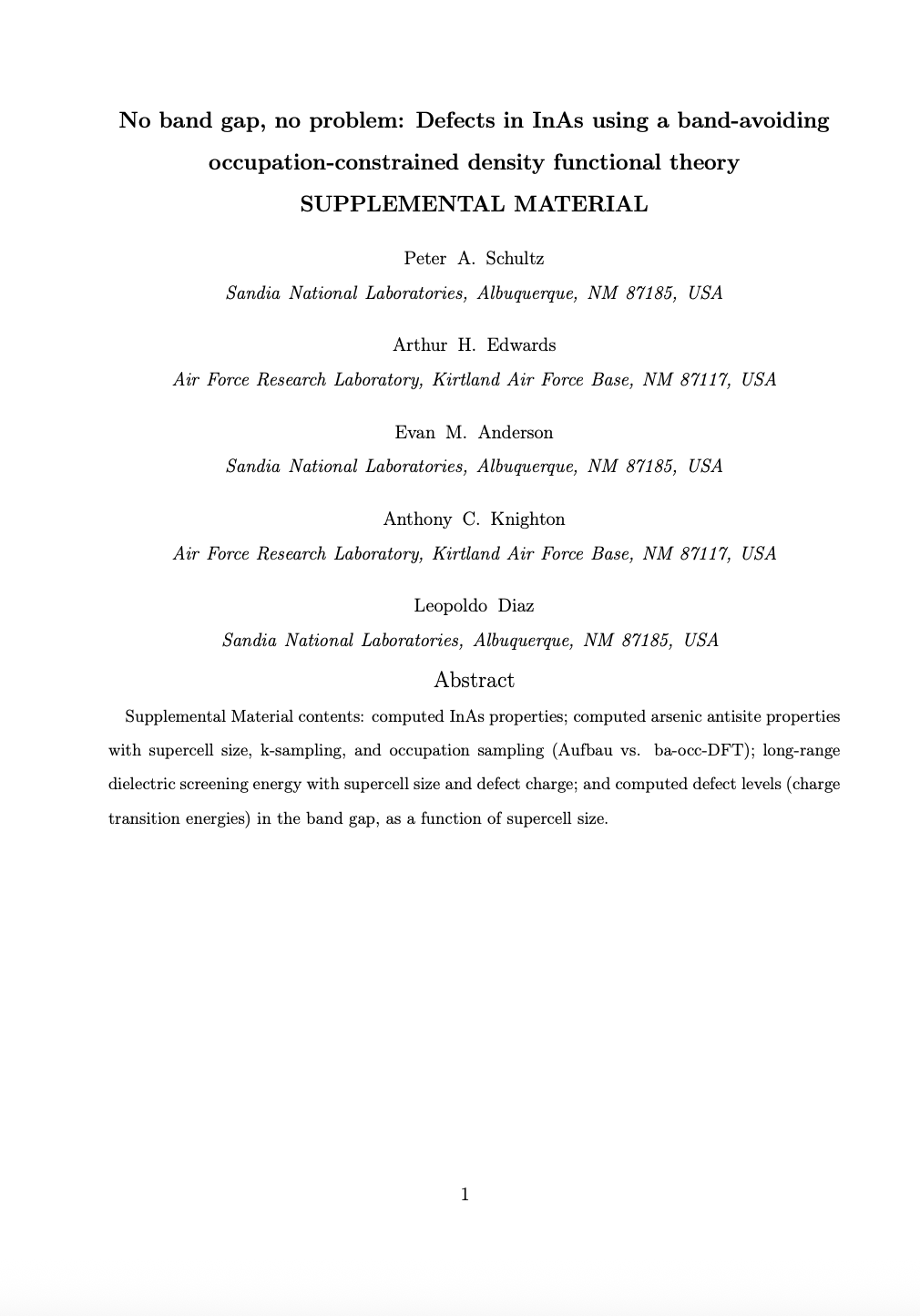}
\end{figure}
\begin{figure} [h!]
    \centering
\includegraphics[width=\textwidth]{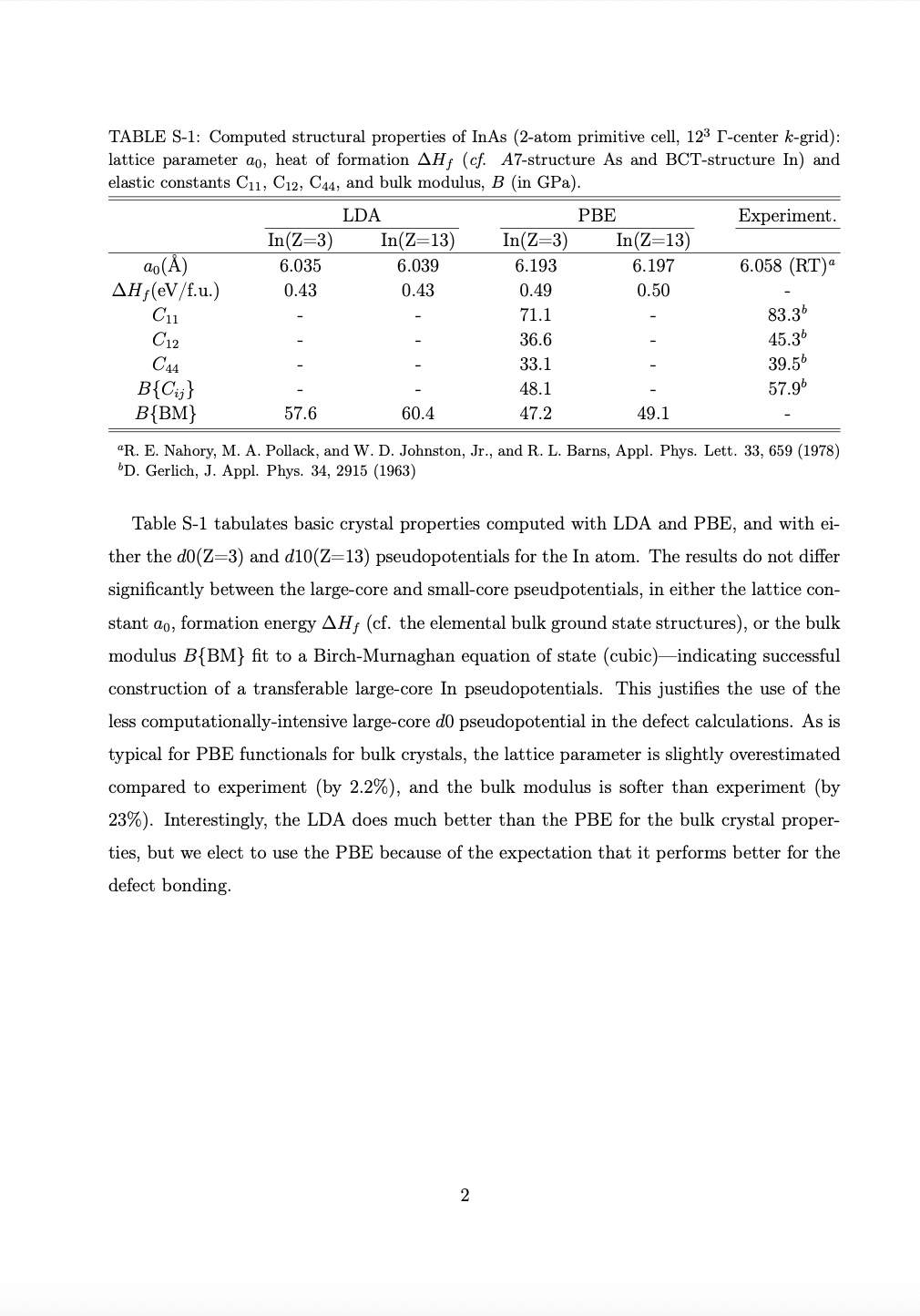}
\end{figure}
\begin{figure} [h!]
    \centering
\includegraphics[width=\textwidth]{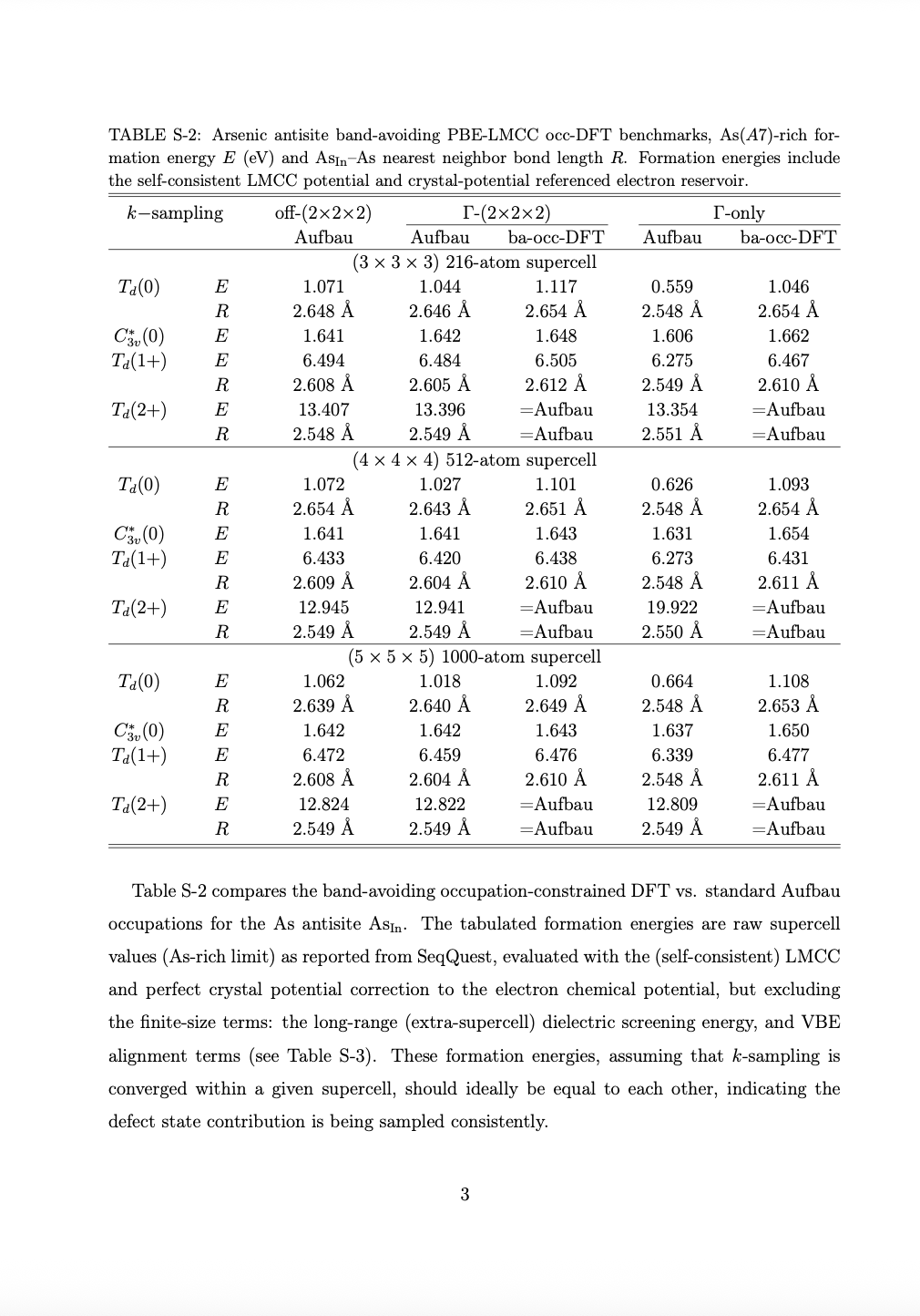}
\end{figure}
\begin{figure} [h!]
    \centering
\includegraphics[width=\textwidth]{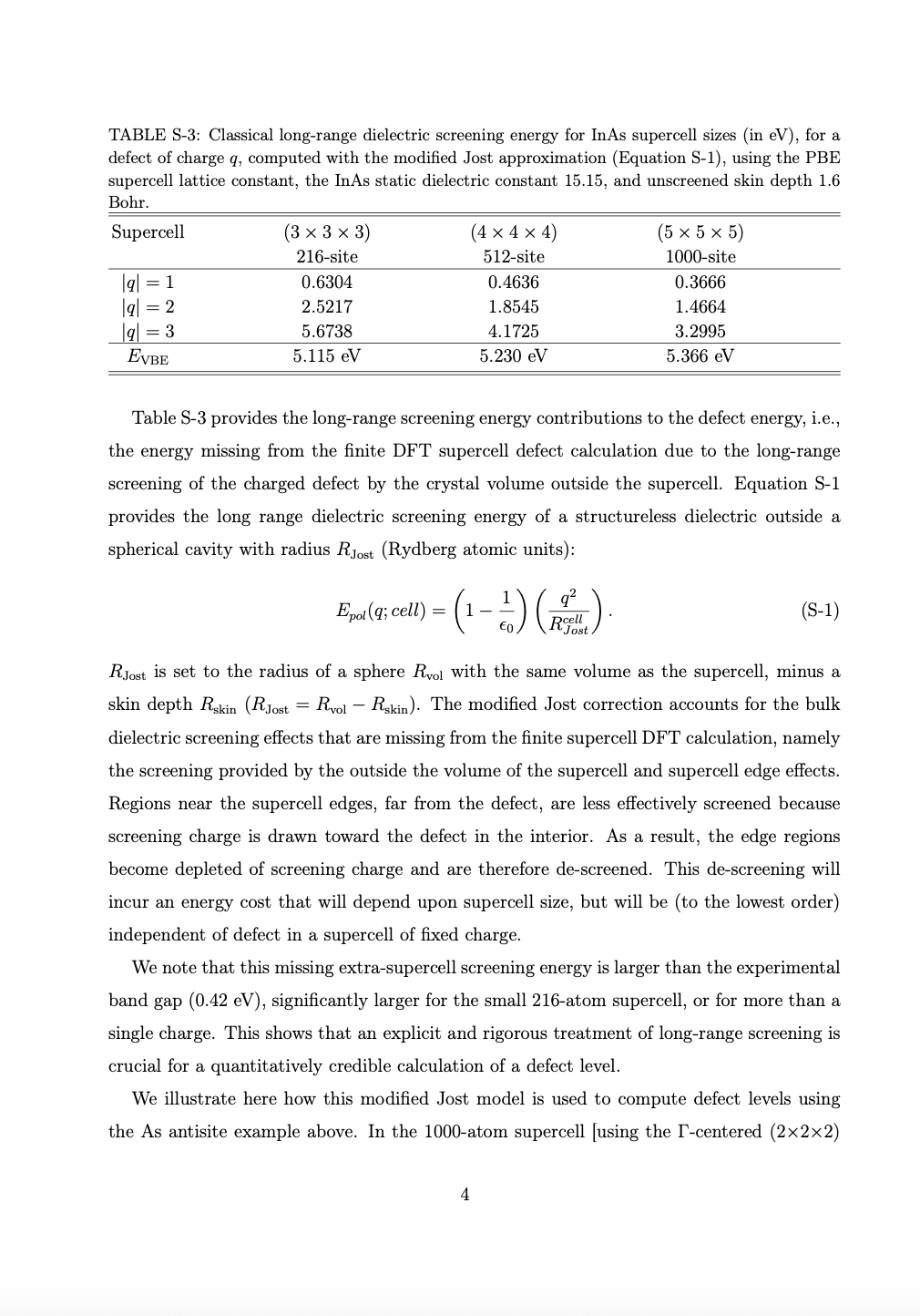}
\end{figure}
\begin{figure} [h!]
    \centering
\includegraphics[width=\textwidth]{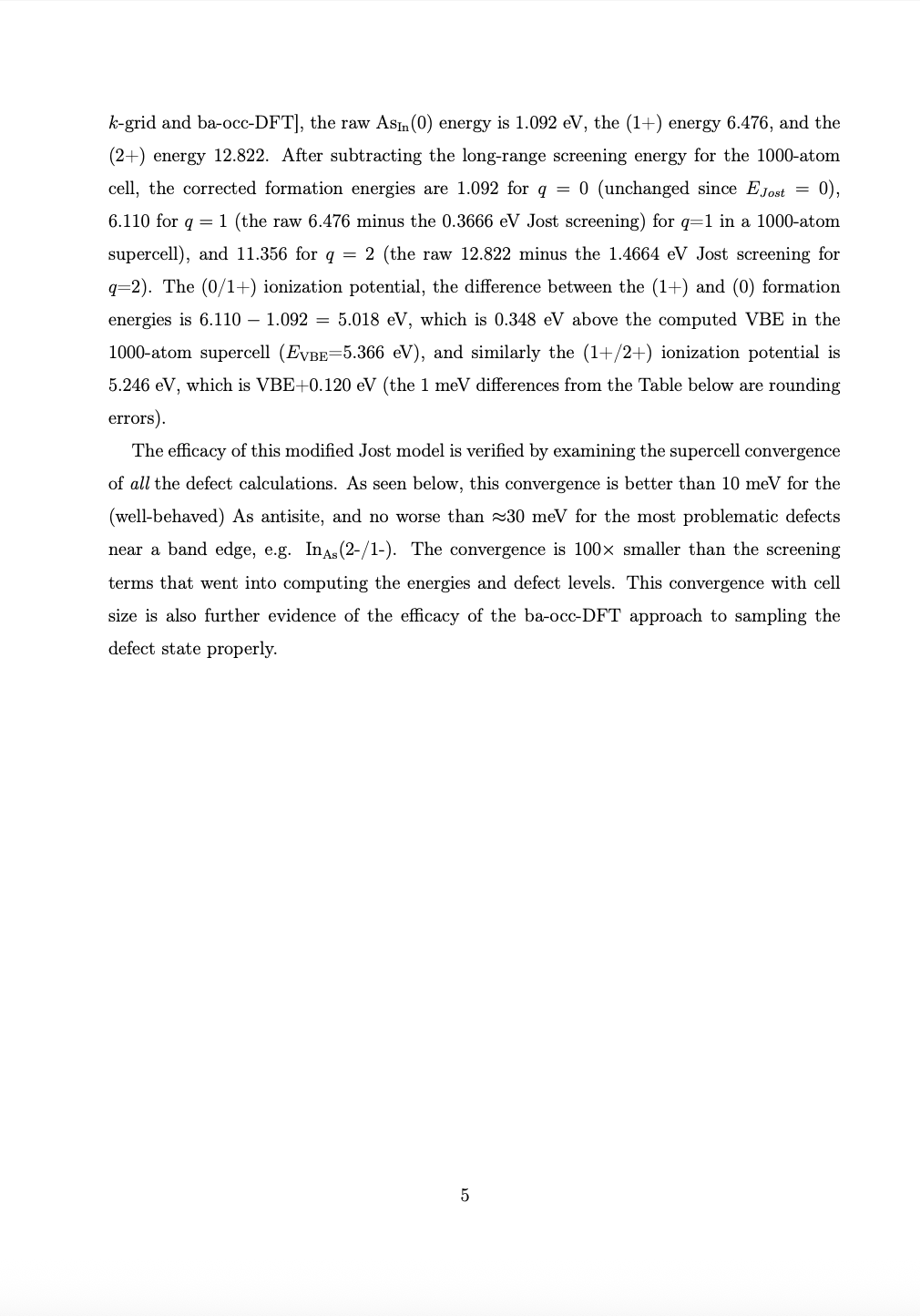}
\end{figure}
\begin{figure} [h!]
    \centering
\includegraphics[width=\textwidth]{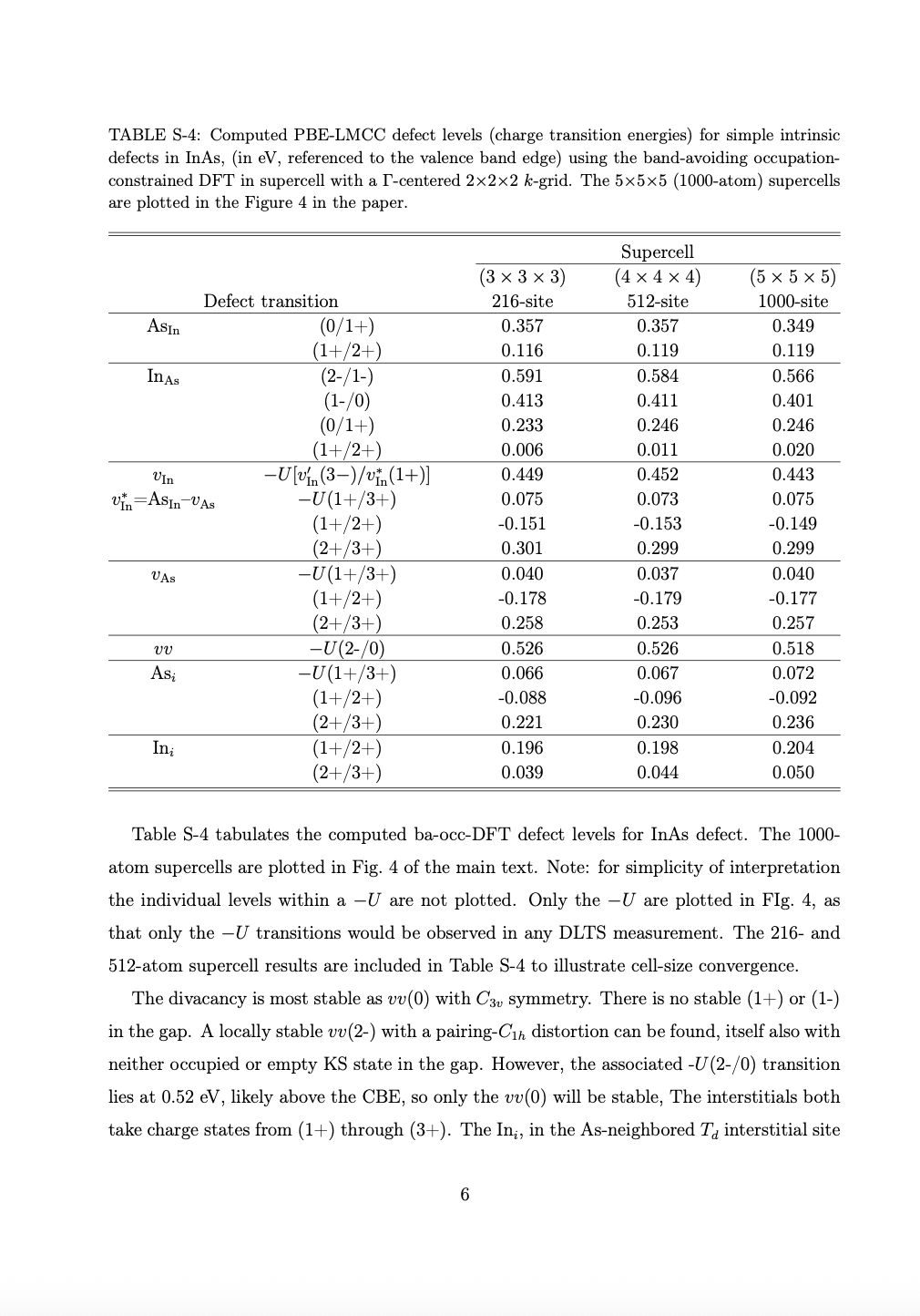}
\end{figure}
\begin{figure} [h!]
    \centering
\includegraphics[width=\textwidth]{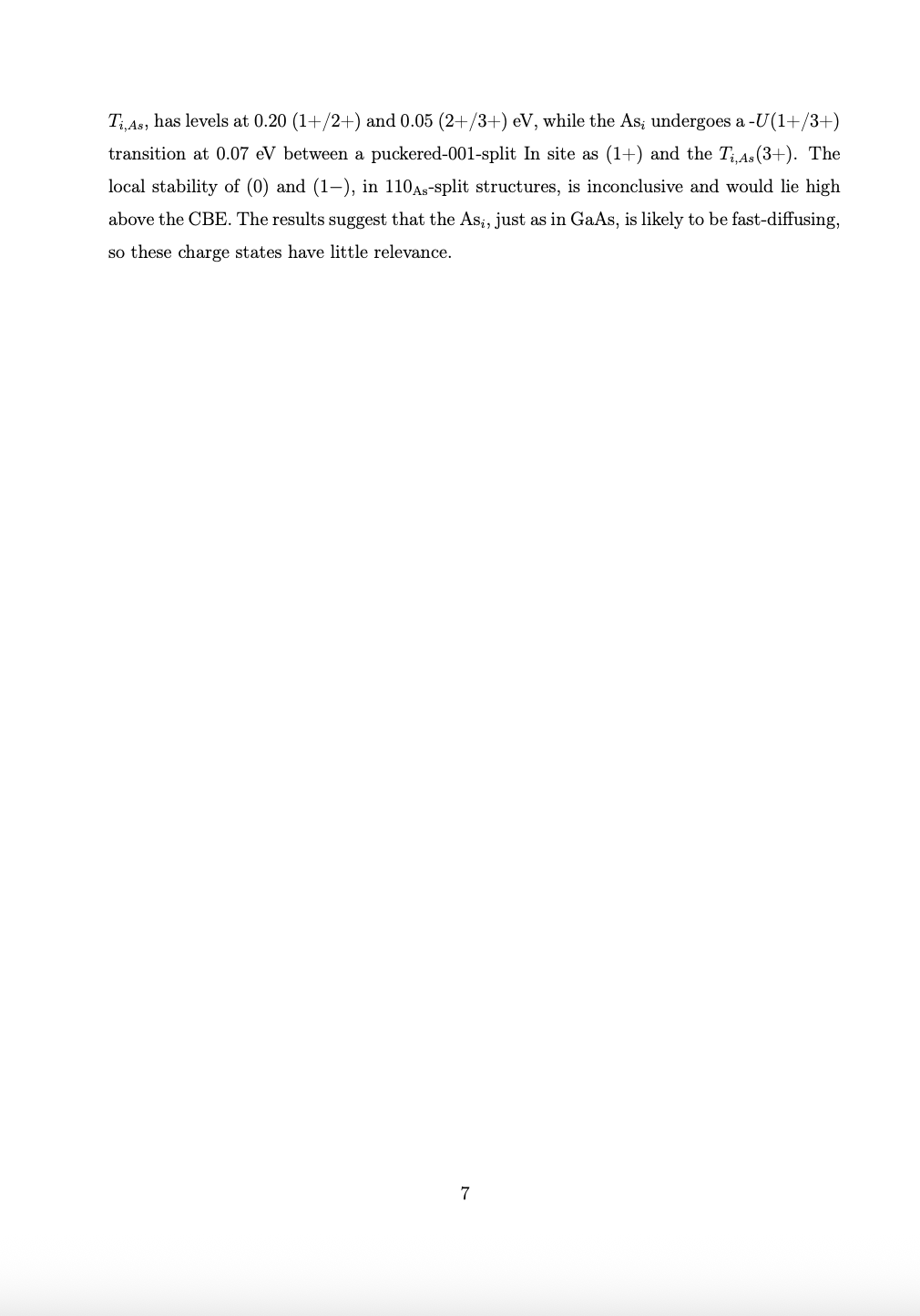}
\end{figure}

\end{document}